# THE THERMAL REGULATION OF GRAVITATIONAL INSTABILITIES IN PROTOPLANETARY DISKS II. EXTENDED SIMULATIONS WITH VARIED COOLING RATES

Running header: Thermal Regulation of GI's in Disks

Annie C. Mejía
Department of Astronomy, University of Washington, Box 351580, Seattle, WA 98195-1580
acmejia@astro.washington.edu

Richard H. Durisen
Department of Astronomy, Indiana University, 727 E. 3rd Street, Bloomington, IN 47405
durisen@astro.indiana.edu

Megan K. Pickett
Department of Chemistry and Physics, Purdue University Calumet, 2200 169th St., Hammond, IN 46323-2094
pickett@astro.calumet.purdue.edu

Kai Cai
Department of Astronomy, Indiana University, 727 E. 3rd Street, Bloomington, IN 47405
kai@astro.indiana.edu

ABSTRACT

In order to investigate mass transport and planet formation by gravitational instabilities (GI's), we have extended our 3-D hydrodynamic simulations of protoplanetary disks from a previous paper. Our goal is to determine the asymptotic behavior of GI's and how it is affected by different constant cooling times. Initially, $R_{disk}$ = 40 AU, $M_{disk}$ = 0.07 $M_\odot$, $M_*$ = 0.5 $M_\odot$, and $Q_{min}$ = 1.8. Sustained cooling, with $t_{cool}$ = 2 orps (outer rotation periods, 1 orp ≈ 250 yrs), drives the disk to instability in ~ 4 orps. This calculation is followed for 23.5 orps. After 12 orps, the disk settles into a quasi-steady state with sustained nonlinear instabilities, an average $Q$ = 1.44 over the outer disk, a well-defined power-law $\Sigma(r)$, and a roughly steady $\dot{M} \approx 5\times10^{-7}$ $M_\odot$ yr$^{-1}$. The transport is driven by global low-order spiral modes. We restart the calculation at 11.2 orps with $t_{cool}$ = 1 and 1/4 orp. The latter case is also run at high azimuthal resolution. We find that shorter cooling times lead to increased $\dot{M}$'s, denser and thinner spiral structures, and more violent dynamic behavior. The asymptotic total internal energy and the azimuthally averaged $Q(r)$ are insensitive to $t_{cool}$. Fragmentation occurs only in the high-resolution $t_{cool}$ = 1/4 orp case; however, none of the fragments survive for even a quarter of an orbit. Ring-like density enhancements appear and grow near the boundary between GI active and inactive regions. We discuss the possible implications of these rings for gas giant planet formation.

Subject headings: Accretion disks – hydrodynamics – planetary systems: formation – planetary systems: protoplanetary disks

1. INTRODUCTION

There are a number of mechanisms that have been considered as possible causes for angular momentum transport in protostellar and protoplanetary disks (see, for example, Morfill, Spruit, & Levy 1993; Adams & Lin 1993; Papaloizou & Lin 1995; Lin & Papaloizou 1996; and Stone et al. 2000), including baroclinic instabilities (Klahr & Bodenheimer 2003 and references therein), magneto-rotational instabilities (Velikhov 1959; Balbus & Hawley 1991, 1997, 2000; Hawley, Gammie, & Balbus 1996; Armitage 1998), and gravitational instabilities, among others. For some mechanisms, inward mass transport can be expressed in terms of an effective turbulent viscosity (Pringle 1981), which assumes that small turbulent (random) motions of the gas mix material from two adjacent, differentially orbiting annuli in the disk. Shakura & Sunyaev (1973) introduced the dimensionless parameter $\alpha$ (< 1) such that the effective viscosity $\nu$ can be written as a function of the gas sound speed $c_s$ and the disk scale height $H$,

$$\nu = \alpha c_s H. \qquad (1)$$

With this prescription, the steady-state mass accretion rate can be written as a function of $\nu$ and the disk surface density $\Sigma$ as follows

$$\dot{M} = 3\pi\nu\Sigma = 3\pi\alpha c_s H\Sigma. \qquad (2)$$

These simple relations do not imply that $\alpha$ is a constant with time or even with position throughout the disk.

As reviewed in Durisen et al. (2001) and Durisen, Mejía, & Pickett (2003), disks become unstable to growth of perturbations due to their self-gravity when the Toomre stability parameter (Toomre 1964), defined as

$$Q = c_s\kappa/\pi G\Sigma, \qquad (3)$$

becomes less than about 1.5 to 1.7. Here, $c_s$ is the sound speed, $\kappa$ is the epicyclic frequency, and $\Sigma$ is the surface density. In disks that undergo such gravitational instabilities (GI's), angular momentum is transferred outward due to torques exerted by high concentrations of mass in the form of trailing spiral arms (Cassen et al. 1981; Larson 1984; Durisen et al. 1986; Adams, Ruden, & Shu 1989; Papaloizou & Savonije 1991; Tomley, Cassen, & Steiman-Cameron 1991; Tomley, Steiman-Cameron, & Cassen 1994; Miyama, Hayashi, & Narita 1984; Laughlin &

Bodenheimer 1994; Laughlin & Różyczka 1996; Bate 1998; Nelson et al. 1998, Nelson, Benz, & Ruzmaikina 2000; Pickett et al. 1998, 2000a, 2003; Yorke & Bodenheimer 1999). The amplitude of GI's, and hence the strength of their influence on the disk, is directly regulated by the balance between cooling and heating processes in the disk. A hot, gravitationally stable disk without sources of dissipative heating will cool until it becomes unstable, and shocks associated with nonlinear growth of the instabilities will generate heat, thus creating feedback between the two processes. It is thought that this balance may keep the disk marginally unstable, with a constant Toomre $Q$ of order unity. This idea was first proposed by Goldreich & Lynden-Bell (1965), studied by Paczyński (1978), subsequently developed by Lin & Pringle (1987, 1990), and tested in numerical simulations by Tomley et al. (1991, 1994) and more recently by Gammie (2001). These simulations confirm that a nearly constant average $Q$ as a function of time can result from the feedback loop (see also Bertin 1997). Moreover, they show that strong cooling, i.e., cooling times of the order of dynamical or orbital times, leads to clump formation by fragmenting the disk quickly before it can react to the gravitational instabilities and heat up (Gammie 2001). Some groups (Lin & Pringle 1987; Laughlin & Bodenheimer 1994; Gammie 2001) have suggested that gravitational instabilities can be approximated by the turbulent viscosity parameter $\alpha$. However, others have argued that GI's are global (i.e., disk-wide) and fully 3-D in character, not local as required by such approach (Laughlin & Różyczka 1996; Balbus & Papaloizou 1999; Pickett et al. 2000a). Recent work by Lodato & Rice (2004) suggests that the degree of locality of GI's may depend on the ratio of the disk to stellar mass. Fragmentation of protoplanetary disks due to GI's is currently of great interest as a possible mechanism for rapid gas giant planet formation (Boss 1997, 1998, 2000, 2001, 2003, 2004; Mayer et al. 2002, 2004).

Many results found in the literature on GI's as a transport mechanism are limited because they are not done in 3-D and because the few simulations evolved for many dynamical times tend to be local, not global. The Tomley et al. calculations represented fairly preliminary work, using a thin disk $N$-body code to mimic hydrodynamics. Gammie's calculations employ full hydrodynamics but are localized to a 2-D shearing box. The present paper is a follow-up to Pickett et al. (2003, hereafter Paper I) in which we extend one of the Paper I grid-based, global 3-D hydrodynamics simulations to almost 24 outer disk orbits. The disk evolves beyond transient behavior and relaxes to a state of equilibrium between cooling and heating for many orbits, revealing its asymptotic response to the thermal processes involved (see also Lodato & Rice 2004). We also investigate how different cooling rates affect the amplitude of GI's, transport of material through the disk, and the fragmentation of the disk into dense substructures.

This paper is organized as follows. In §2, we briefly describe the three-dimensional hydrodynamics code and its internal energy equation. The simulations and their results are presented in §3. In §4 we discuss the implications of these simulations and compare them with other similar works. Summary and conclusions are presented in §5.

## 2. THREE-DIMENSIONAL HYDRODYNAMICS

The three-dimensional hydrodynamics code with self-gravity is the same as that used in Paper I and is described in detail in Pickett (1995), Pickett et al. (1998, 2000a), and Mejía (2004). To summarize, the hydrodynamics code solves Poisson's equation, an ideal gas equation of state, and the equations of hydrodynamics (Yang 1992) in conservative form on a uniform cylindrical grid ($r,\phi,z$). The code computes the source and flux terms (Norman & Winkler 1986) separately in an explicit, second-order time integration scheme (van Albada, van Leer, & Roberts 1982; Christodoulou 1991; Yang 1992). Although the code is identical to the one used in Paper I, a few details are repeated here for the reader's convenience.

2.1. Boundary Conditions

The 3-D hydrodynamics code assumes mirror symmetry across the equatorial plane, so the equatorial plane of the disk is situated along the lowest $z$ zone. The cells at the edge of the grid (highest $z$ and $r$) assume outflowing boundary conditions, i.e., the momentum densities in these cells are always positive and the material flows out of the grid. Since the disk models do not include the star physically, but only its gravitational potential, it makes sense to stop hydrodynamic calculations somewhere inside the inner radius of the disk. The initial disk model used for the simulations in this paper (Figure 3 in Paper I, top panel) has its inner edge at radial zone $J = 16$, so calculations are not performed inside $J = 13$, where an inner outflow boundary is placed. All the material crossing this inner boundary is considered to have accreted onto the star, therefore its potential is added to that of the star by assuming it falls to the grid's origin, i.e., as a point source potential. The momentum density of accreted material is not added to the central star, which is held fixed in place at the coordinate origin throughout the calculation.

2.2. The Internal Energy Equation

If an ideal gas equation of state $P = (\gamma - 1)\varepsilon$ is assumed, where the gas pressure $P$ and the internal energy density $\varepsilon$ are related by $\gamma = 5/3$ for a monatomic ideal gas, then the energy equation can be written in conservative form in terms of $\varepsilon^{1/\gamma}$ (Williams 1988; Pickett et al. 2000a)

$$\frac{\partial(\varepsilon^{1/\gamma})}{\partial t} + \nabla \cdot (\varepsilon^{1/\gamma}\mathbf{v}) = \frac{1}{\gamma}\varepsilon^{1/\gamma-1}(\Gamma - \Lambda). \tag{4}$$

In the simulations presented here, the heating term $\Gamma$ includes only the effects of shock heating by artificial viscosity, using the von Neumann & Richtmeyer scheme (Norman & Winkler 1986). The cooling term $\Lambda$ is applied throughout the entire volume of the disk, assuming $t_{cool} = \varepsilon/\Lambda =$ constant. The terms $\Gamma$ and $\Lambda$ are nonzero only when $\rho/\rho_{min} \geq 10^5$, because cells with lower density are not considered part of the disk structure, but of its diffuse surrounding. The minimum density that any grid cell can contain, $\rho_{min}$, is set at the beginning of every simulation to be $10^{-12}$ times $\rho_{max}$, the density at the center of the (now missing) star of the original 2-D star/disk model (see §2.3 in Paper I). The low value assigned to $\rho_{min}$ guarantees that a negligible amount of mass is added to the computational grid every time a cell's density is reset to the limiting value. For the 2-D initial disk model used in the simulations, the maximum initial disk density is $8.8 \times 10^{-4} \rho_{max}$, so the internal energy equation is calculated across 3.9 orders of magnitude in density.

3. THE SIMULATIONS

3.1. Initial Model

In contrast to Paper I, the initial axisymmetric model is scaled to have a central star of 0.50 $M_\odot$ and a nearly Keplerian disk of 0.070 $M_\odot$, which extends from 2.3 to 40.0 AU. The disk's surface density distribution is roughly a power law, with $\Sigma \propto r^{-1/2}$. Figure 1 illustrates the meridional cross-section and the radial run of several physical parameters at the equatorial plane. The initial disk is marginally stable, with a minimum $Q \approx 1.8$ near 30 AU. The number density varies from $\approx 10^{13}$ to $10^8$ cm$^{-3}$. Assuming ideal gas and a mean molecular weight of 2.717 proton masses given by a mean molecular weight table from P. D'Alessio (1996) yields midplane temperatures ranging from about 330 K at r = 3.2 AU to a forced minimum of 3 K.

The initial model is placed in the 3-D hydrodynamics code with an initial grid of size $(r,\phi,z) = (256,128,32)$, where one radial (or vertical) grid zone is exactly 1/6 AU. The disk extends from radial zone 16 to 242 and the largest vertical thickness is 27 cells from the equatorial plane. A random cell-to-cell density perturbation of amplitude $|\Delta\rho/\rho| = 10^{-4}$ is applied to the initial model. The volumetric cooling and shock heating are only computed in cells with number density larger than $1.05 \times 10^9$ cm$^{-3}$, which corresponds to the mass density limit described in §2.2.

3.2. Four Cases

The cooling in these simulations is not meant to be particularly realistic, but it does allow for controlled experiments by using a constant (in time and space) cooling time $t_{cool}$ throughout the disk. Simulations we have done using realistic radiative cooling, which will be presented in a forthcoming paper (see also Mejía 2004), show $t_{cool}$ does become roughly constant. So our assumption of a constant $t_{cool}$ is probably more realistic than the $t_{cool}\Omega$ = constant assumed by other researchers (e.g., Rice et al. 2003; Lodato & Rice 2004), where $\Omega$ is the local angular speed. Three simulations, one with $t_{cool}$ = 2 orps (see below), another with $t_{cool}$ = 1 orp, and a third with $t_{cool}$ = 1/4 orp are conducted in order to examine the asymptotic behavior of GI's for different cooling times in disks that evolve under otherwise identical conditions, as a follow-up to the work by Tomley et al. (1991, 1994). A fourth simulation, also using $t_{cool}$ = 1/4 orp but with quadrupled azimuthal resolution, is run to study how much the characteristics of the evolution depend on resolution (Paper I) and to determine whether the disks comply with Gammie's (2001) fragmentation criterion, namely, that dense clumps will form if $t_{cool}\Omega \leq 3$.

The main time unit used in this paper is the orp, defined as the "outer" rotation period of the initial disk model at radial zone 200 (33 AU), chosen arbitrarily. In these simulations, 1 orp ≈ 250 yr. The $t_{cool}$ = 2 orps (or just $t_{cool}$ = 2 for short) calculation starts with the axisymmetric disk model described in the previous section and evolves for a total of 23.5 orps, or about 5,875 yr. The computational grid is doubled in size in the vertical and radial directions, $(r,\phi,z)$ = (512,128,64), to accommodate the disk after the expansion begins, between 5 and 6 orps (Figure 2). The rest of the simulation is carried out on the large grid, and so are the entire $t_{cool}$ = 1 and $t_{cool}$ = 1/4 runs. Since we are mainly interested in studying the influence of $t_{cool}$ on the quasi-steady GI behavior, we start the shorter $t_{cool}$ simulations from the $t_{cool}$ = 2 disk after 11.2 orps of evolution, when the disk transients have disappeared. The high azimuthal resolution calculation (or High-res, for short) is mostly a repeat of $t_{cool}$ = 1/4, performed here with four times the azimuthal resolution but on a grid (256,512,32) truncated in $r$ and $z$. The $t_{cool}$ = 1 disk evolves for a little over 6 orps with the new cooling; it is stopped at 18 orps of total evolution time. The $t_{cool}$ = 1/4 and High-res disks evolve less than 3 orps, ending at 14 orps of total evolution time. Table 1 lists several parameters for the initial model ($t$ = 0) and for the $t_{cool}$ = 2 disk at 11.2 orps at 10, 30, and 50 AU from the origin.

TABLE 1
Characteristics of the $t_{cool}$ = 2 Disk at 0 and 11.2 orps

| $M_*$ ($M_\odot$) | $M_{disk}$ ($M_\odot$) | $t$ (orps) | $R_{disk}$ (AU) | Grid size $(r,\phi,z)$ | $r$ (AU) | $T$ (K) | $Q$ | $n$ (cm$^{-3}$) | $\Sigma$ (g cm$^{-2}$) |
|---|---|---|---|---|---|---|---|---|---|
| 0.5 | 0.07 | 0 | 40.0 | 256,128,32 | 10 | 102.5 | 7.4 | 1.5(12) | 207.9 |
| | | | | | 30 | 45.9 | 1.8 | 4.5(11) | 141.6 |
| | | | | | 50 | … | … | … | … |
| | | 11.2 | 57.3 | 512,128,64 | 10 | 243.4 | 2.5 | 5.6(12) | 286.2 |
| | | | | | 30 | 57.3 | 1.4 | 6.4(11) | 60.4 |
| | | | | | 50 | 7.5 | 3.4 | 3.0(10) | 9.4 |

3.3. The $t_{cool}$ = 2 Simulation

Figure 2 (upper section) shows number density contours of the equatorial plane of the $t_{cool}$ = 2 disk every two orps for the entire evolution. The disk preserves axisymmetry until about 2 orps, but it shrinks slightly in the radial direction because it cools continuously. It first becomes unstable at 3 orps with minimum $Q < 1$. After 4 orps, the instabilities reach nonlinear amplitude, and the resulting nonaxisymmetries become apparent in the equatorial plane. Soon afterwards, large global spiral disturbances appear, dominated initially by a four-armed mode, and the disk expands violently (Figure 2, bottom section). The spiral structure of the disk is strongest at 6 orps, which is also approximately when most of the mass transport within the disk occurs. This initial "burst" of GI activity in a discrete mode is reminiscent of low-order mode instabilities found in polytropic stars and disks (Durisen et al. 1986; Pickett, Durisen, & Davis 1996; Imamura, Durisen & Pickett 2000). The strong spiral pattern dies quickly due to shock heating, and some of the nonaxisymmetric structure is lost for another 6 orps. After 12 orps, the disk develops new, thinner and denser spiral arms, and the subsequent global structure is quasi-steady, i.e., the detailed structure varies on a dynamic time scale, but average quantities change steadily but slowly.

3.3.1. Mass and Density Distribution

The overall structure of the disk changes drastically after the instabilities appear. There are both inward and outward mass transport zones, separated by a definite boundary. Most of the inward mass transport happens between 10 and 29 AU, with peak transport rates of $10^{-5}$ $M_\odot$ $yr^{-1}$ during the burst between 5 and 6 orps, while the outer disk expands at a similar rate. After 12 orps, the average accretion rates are of the order of $5\times10^{-7}$ $M_\odot$ $yr^{-1}$ between 10 and 29 AU with peaks at $10^{-6}$ $M_\odot$ $yr^{-1}$. Outside 29 AU, mass flows outward with similar average rates as the disk expands. Figure 3 shows mass transport rates versus $r$ averaged over 12 to 18 orps and 18 to 23.5 orps, as computed by comparing the mass interior to a cylinder at $r$ at the end times of the intervals. Also shown for comparison are the extreme fluctuations that occur when the same procedure is used over much shorter time intervals between 12 and 23.5 orps. The average mass transport rate in the 10 to 29 AU region remains reasonably steady over the entire asymptotic phase of the simulation, even though the instantaneous fluctuations in $\dot{M}$ can be extremely large and even reverse sign. As discussed in more detail below, there is redistribution of mass inside 10 AU as suggested by the appearance of nonaxisymmetric patterns, but no steady mass inflow.

The initial surface density distribution $\Sigma(r) \propto r^{-1/2}$ is lost as the disk becomes nonaxisymmetric. During the onset of the instabilities it becomes very steep in the outer regions, but after 12 orps a least squares fit shows that the azimuthally averaged $\Sigma(r)$ follows a power law $\propto r^{-2.50}$ (with a correlation coefficient $R^2$ of 0.94) between 14 and 50 AU. Over the last 12 orps, the average disk edge expands at approximately 1.4 AU orp$^{-1}$ or 0.006 AU yr$^{-1}$.

The most unexpected but exciting feature of this simulation is the steady growth of a series of concentric rings in the inner disk centered around 3.5, 6.7, and 10.2 AU (Figure 4), which form in sequence from innermost to outermost. The mass of the rings is approximately 1.2, 3.0, and 8.8 $M_J$, respectively, with number densities as high as $2.4 \times 10^{13}$ cm$^{-3}$. For comparison, the original masses at these locations are 1, 2, and 5 $M_J$, measured within the same radii as the rings. The disk also develops a number of arc-like features along the spiral arms between 14 and 17 AU. Although none grow into bound permanent objects, the presence of such transient arc-like overdensities is a consistent feature of the entire evolution from 12 to 23.5 orps. Some do in fact persist for several orbits. There is as much as 9.7 $M_J$ between 14 and 17 AU at the end of the evolution, versus 5.5 $M_J$ originally.

3.3.2. Fourier Analysis

As a standard diagnostic technique in our studies of gravitational instabilities (e.g., Pickett et al. 1998, 2000a), we regularly store density images of the equatorial plane and Fourier decompose the azimuthal $\delta\rho/\rho_0$ in $\cos(m\varphi + \varphi_m)$. Here, $\rho_0$ is the azimuthally averaged density, $\delta\rho$ the nonaxisymmetric deviation from $\rho_0$, and $\varphi_m$ is the instantaneous phase of the $m$-armed fluctuation. In addition to determining Fourier component density amplitudes $\delta\rho_m$, we can detect any pattern frequencies $\Omega_{pat}$ for each $m$ that are coherent over a range of radii by analyzing the time series for the $\varphi_m$ at each $r$. We can then identify the Corotation Radii (CR's) ($\Omega = \Omega_{pat}$) and inner and outer Lindblad Resonances ($\Omega = \Omega_{pat} \pm \kappa/m$) (OLR's and ILR's) associated with strong patterns. For $t_{cool} = 2$, we have analyzed Fourier component amplitudes and time series information for $m = 1$ to 6 from the time intervals 11 to 18 orps and 18 to 23.5 orps in order to characterize the complex nonlinear structure present in the asymptotic state.

The disk has numerous spiral wave patterns ($m$'s) appearing in the outer low-$Q$ part of the disk, typically confined between their corresponding inner and outer Lindblad resonances. Results are illustrated in Figure 5 for $m = 2$. Notice the pervasiveness of coherent patterns in the outer disk. Plots for $m = 1$ and $m > 2$ are similar except that the density of apparently discrete modes increases with $m$ in a manner consistent with strong nonlinear mode coupling, and distinct $m = 1$ patterns do not tend to extend over a range of radii except in the outer disk. Although some patterns are isolated in period or radius, the wide array of periods detected suggests pervasive nonlinear interaction of waves throughout the low-$Q$ disk, what Gammie (2001) calls "gravitoturbulence." There appear to be two dominating coherent $m = 2$ patterns in the midplane between 13 and 37 AU (Figure 5), with CR's ≈ 26 to 29 AU, variable pattern periods ≈ 0.7 to 0.8 orps, and strength typically $\delta\rho_2/\rho \approx 0.3$, but with considerable spatial and temporal variation. The CR's are at roughly the same radius that separates inward from outward mass transport, indicating that these $m = 2$ patterns dominate the torques that redistribute mass. Thus, the mass transport is dominated by global, not local modes. Other patterns ($m = 1, 3, 4$) can, at times, in various locations, be as strong or stronger than $m = 2$, but higher order patterns ($m = 5, 6$) are generally weaker. A few ILR's appear in regions between the rings, e.g., the $m = 2$ disturbance with period ≈ 0.29 orps in Figure 5. The possible role of resonances in ring formation will be discussed in § 4.3.

3.3.3. Heating and Cooling

By our assumption that $t_{cool} = \varepsilon/\Lambda = 2$ orps, the disk is instantaneously losing energy at a rate equal to all its internal energy every 2 orps. Figure 6 shows the total internal energy of the disk as a function of time. During the first part of the evolution while the disk is axisymmetric, the cooling causes the total internal energy to decrease steadily because shocks have not yet developed, which in turns drives the initial slow contraction of the disk. However, the rate of change in internal energy is not equal to the cooling rate because there is some heating caused by vertical and radial compression. Once the disk reaches instability after a few orps of continuous cooling, the development of dense spiral structures with non-Keplerian motions produces substantial shock heating. Some of this heating is due to radial fall back of the strong material spiral arms that are ejected in the initial burst (bottom of Figure 2). The total internal energy $U_{tot}$ increases abruptly and it is approximately maintained for a couple of orbits. Analysis of the $Q(r)$ distribution between 4.5 and 7 orps shows that the added energy makes the disk stable, the strong spirals disappear, and the energy input decreases. The continuous cooling slowly brings the disk back to instability again by about 12 orps. This time the onset of instability is gentle and produces thin and moderately dense arms in the mid-disk, which survive the rest of the run. Subsequently, cooling is roughly balanced by heating, and the total internal energy approaches an asymptotic value close to $10^{41}$ erg, a little less than a third of the initial value. The luminosity of the disk follows the same curve as the internal energy in Figure 6, and the scale is indicated on the second vertical axis in solar luminosities. It is important to remember that most of the luminosity of a disk like this would come from regions closer to the star, where the gas and dust are the hottest. The region inside 2.3 AU is not included in this simulation, which accounts for the low total luminosity.

In addition to the internal energy, the overall structure of the disk, the Toomre $Q(r)$, surface density distribution, and other disk properties achieve approximately asymptotic states after 12 orps. The disk settles to an average $Q = 1.44$ with a standard deviation $\sigma = 0.24$ between 12 and 40 AU by 12 orps. However, because the disk is in a strongly variable nonlinear dynamic state, its properties fluctuate on a dynamic time scale even when the averages are mostly constant. Figure 7 shows the distribution of $Q$ versus radius at various times during the evolution. Outside 50 AU the surface density of the disk is low and so $Q$ is higher. The disk inside 10 AU remains too hot to be GI unstable, except possibly in the dense rings where $Q$ can be considerably lower.

Figure 8 illustrates disk properties (midplane number density, heating time, temperature, and effective temperature) at 6.1 orps (top panel) and at the end (bottom panel) of the evolution. The number density contours are as in Figure 2. The midplane heating time maps show that the heating caused by shocks, with the shortest heating times as low as 0.05 orps at 6.1 orps, occurs right next to the spiral arms, where compression is strongest. Shocks occur on the trailing side of the spiral arms inside the corotation radius, where the gas in orbit moves faster than the gas in the spiral arms. Outside the CR, the period of the spiral arms is shorter than the orbital period, so the shocks occur on the leading side (Paper I). At the end of the run, shock heating is not as strong because the disk is more quiescent. The typical heating times are of the order of several tens of orps or even several hundred orps in some parts of the disk outside the shock regions.

The midplane temperatures near the end of the calculation range between 300 and 3 K, not too different from the initial temperatures. The highest temperatures are exclusively in the innermost part of the disk (≤ 5 AU), quickly decreasing to tens of Kelvin in the rest of the equatorial plane. The effective temperature $T_{eff}$ map, which approximates what an observer would see if the disk were face on in the sky, shows $T_{eff}$ ranging between 40 and 1.5 K. Effective temperatures are obtained by assuming that the integral of the volumetric cooling rate from midplane to disk surface along the vertical direction equals the surface flux from the column. $T_{eff}$ can be higher than the midplane temperatures in some parts of the disk, especially in the outer disk. This is partly due to the temperatures in the disk being higher above the midplane. An azimuthal average of the effective temperatures vs. radius at the end of the evolution yields $T_{eff} \propto r^{-0.71}$ between 3 and 50 AU (with correlation coefficient $R^2 = 0.87$). If, however, the fit is done in the nearly constant $Q$ region, between 12 and 40 AU, the effective temperature follows a steeper $r^{-0.84}$ ($R^2 = 0.95$) power law.

The vertical structure of the disk at 6.1 and 23.5 orps out to 50 AU is shown in the maps of Figure 9, where the vertical scale (5 AU) is exaggerated by a factor of 1.5. The top two groups of maps show the disk right after the first expansion. The left group of maps shows the slice of the disk that corresponds to the 3 o'clock position in the equatorial maps of Figure 8. The right group shows the average of all the azimuthal slices. Notice the significant departure from the vertical structure shown in Figure 1 as the disk first becomes gravitationally unstable. More of the disk is in the high end of the temperature scale, hundreds of Kelvin, at 6 orps than at the end of the run. As discussed by Pickett et al. (2000a), Figure 9 shows that shock heating is strongest in the upper layers of the disk, both in the meridional slice and in the azimuthal average. As a result, the highest temperatures are systematically above the midplane at both times. Thin disk treatments of GI's miss this important effect. Midplane temperatures are not representative of the real 3-D temperature distribution in the disk. When meridional slice velocity vectors are superimposed on the artificial viscosity heating time maps, the shortest heating times occur where arrowheads meet, as expected. Strong spiral shocks seem to be associated with strong upward and downward motions, and we are currently investigating whether shocks in stratified disks exhibit some features of hydraulic jumps (Mantos & Cox 1998).

## 3.4. The $t_{cool} = 1$ and $t_{cool} = 1/4$ Simulations

The initial conditions for the $t_{cool} = 1$ and $1/4$ simulations are the $t_{cool} = 2$ disk at 11.2 orps, and they are both run for about 6 $t_{cool}$, which appears to be sufficient time for the system to relax into a new quasi-equilibrium. The $t_{cool} = 1$ simulation ends at about 18 orps of total evolution time, while the $t_{cool} = 1/4$ ends at about 14 orps. In general, these simulations are more dynamic than the $t_{cool} = 2$ run and the quasi-steady spiral structure becomes more sharply defined and denser as $t_{cool}$ decreases.

### 3.4.1. Mass and Density Distribution

The mass transport rates within the disks are higher than in the $t_{cool} = 2$ case, with an average rate of approximately $10^{-6}$ $M_\odot$ yr$^{-1}$ for $t_{cool} = 1$ and $4 \times 10^{-6}$ $M_\odot$ yr$^{-1}$ for $t_{cool} = 1/4$ (Figure 10) over the 10 to 27 AU region. The point of separation between inward and outward transport is 26 AU and 27.5 AU for $t_{cool} = 1$ and $1/4$, respectively. As in the previous case, there is no significant mass transport in the innermost part of the disks ($\leq 7$ AU). The amount of mass leaving the grid out the upper and outer boundaries during the simulations is also negligible. The $t_{cool} = 1$ disk surface density follows a $r^{-2.47}$ power law between 10 and 60 AU ($R^2 = 0.93$). For $t_{cool} = 1/4$, $\Sigma \propto r^{-2.48}$ ($R^2 = 0.84$) between 14 and 60 AU. Both power laws are measured at the end of the evolution. The $t_{cool} = 1$ disk expands twice as fast as $t_{cool} = 2$, about 2.7 AU orp$^{-1}$ or 0.01 AU yr$^{-1}$. Since the disk in the $t_{cool} = 1/4$ simulation reacted so violently to the enhanced cooling, no expansion rate could be reliably measured.

The high-density features observed in the $t_{cool} = 2$ disk become even more pronounced in these shorter cooling time runs. The dense rings in $t_{cool} = 1$ grow at the same radii as in the $t_{cool} = 2$ disk, but the masses at the end of the run are higher: 1.5 $M_J$ for the 3.5 AU ring, 4 $M_J$ for the 6.7 AU ring, 8.8 $M_J$ for the 10 AU ring, and 9.5 $M_J$ for the $\sim 15$ AU region. Mass accumulates in the rings about 2 times faster than in the previous calculation. The highest densities at the end of the run are $4.0 \times 10^{13}$ cm$^{-3}$ within the rings, about 20 times denser than the initial ($t = 0$) density at the same location.

Two of the dense rings observed in longer $t_{cool}$ simulations are present in $t_{cool} = 1/4$. At 14 orps, the 3.5 AU ring and the 6.7 AU ring have already formed but contain more mass than the rings of the other two simulations at the same evolutionary time. Another ring appears at 4.3 AU, which seems to have taken the mass from the 5.2 AU region where now there is an annular "void." The 10 AU ring has not formed yet. The region between 8 and 13 AU is dense but still very dynamic, and the spiral arms penetrate all the way into it. At 14 orps, the masses of the rings are 1.2 $M_J$ for each of the two innermost rings and 5.8 $M_J$ for the 6.7 AU ring, already greater at 14 orps than it is at 18 orps for $t_{cool}$ = 1. The highest density at the end of this run is $7.7 \times 10^{13}$ cm$^{-3}$ at the 6.7 AU ring, almost 30 times the initial density at the same location.

3.4.2. Fourier Analysis

In general, all the patterns of $t_{cool}$ = 1 have higher amplitudes in $\delta\rho/\rho$ than $t_{cool}$ = 2 and tend to propagate somewhat outside the Lindblad resonances. The $m$ = 2 (two-armed) patterns are dominant in this simulation as well. The radial distribution of $\delta\rho_2/\rho$ is more uniform, and the amplitude is almost twice as large ($\approx$ 0.5 to 0.8 over 13 to 50 AU) as in the $t_{cool}$ = 2 evolution. The discrete periods associated with $m$ = 2 CR's near 26 to 29 AU, though present, are not as well defined. Unlike the $t_{cool}$ = 1 and 2 calculations, there are no dominant patterns in $t_{cool}$ = 1/4. The $\delta\rho/\rho$ amplitudes appear highly saturated ($\approx$ unity for $m$ = 2 and hitting unity at many radii for $m$ = 3 to 6) over much of the disk. Patterns extend well outside the Lindblad resonances, suggesting the presence of strong waves and oscillations other than spiral density waves. Wave activity in these disks becomes more complex and nonlinear as $t_{cool}$ decreases.

3.4.3. Heating and Cooling

As shown in Figure 11, at 11.2 orps, the total internal energy of both $t_{cool}$ = 1 and $t_{cool}$ = 1/4 is that of the $t_{cool}$ = 2 disk. The energy immediately drops because of the shorter cooling times. Denser and more dynamic spirals appear, which cause shock heating to become stronger. The result is a return of the total internal energy to close to its original value. The noise in Figure 11 makes it uncertain whether the $t_{cool}$ = 1/4 simulation has evolved far enough for the internal energy to have achieved its asymptotic value. Nonetheless, it seems that it will be the same or only a slightly lower value than the final $U_{tot}$ for the $t_{cool}$ = 1 and 2 disks.

The Toomre $Q$ as a function of radius for $t_{cool} = 1$ and $1/4$ is very similar to that of the $t_{cool} = 2$ disk, averaging 1.43 ($\sigma = 0.23$) and 1.50 ($\sigma = 0.37$), respectively, between 12 and 40 AU. However, the distribution of $Q$ for $t_{cool} = 1/4$ is much less smooth. In fact, $Q$ varies between 0.7 and 2.3 over $r = 12$ to 40 AU. In the ring region (< 10 AU), the dips and peaks are more pronounced, as expected, due to the increase in density within the rings.

The effective temperatures in $t_{cool} = 1$ are increased over those in $t_{cool} = 2$ by 20%, as expected ($T_{eff} \propto L^{1/4} \propto t_{cool}^{-1/4}$), while $t_{cool} = 1/4$ shows 50% higher effective temperatures. Power-law fits to $T_{eff}(r)$ yield $T_{eff} \propto r^{-0.73}$ ($R^2 = 0.91$) and $T_{eff} \propto r^{-0.65}$ ($R^2 = 0.86$) between 3 and 50 AU for $t_{cool} = 1$ and $t_{cool} = 1/4$, respectively. In the approximately constant $Q$ region, 12 to 40 AU, $T_{eff} \propto r^{-0.94}$ ($R^2 = 0.96$) and $T_{eff} \propto r^{-0.82}$ ($R^2 = 0.83$) for the same simulations, respectively. The vertical structure of both low $t_{cool}$ cases is not too different from the $t_{cool} = 2$ disk; however, they are, on average, vertically thicker than the latter. The temperature ranges are also similar, and, again, higher temperatures are typically seen above the disk midplane.

3.5. Comparisons

Figure 12 illustrates how decreasing cooling time changes disk structure. Remember that densities lower than $\approx 10^9$ cm$^{-3}$ are not cooled, so the diffuse material in the outer grid maintains more or less the same spatial distribution until it is affected by the denser disk inside. At 12 orps, the dense regions of the disk follow the same spiral patterns, but the spirals become more concentrated (denser and thinner) with shorter cooling times. In addition, the spiral pattern penetrates deeper into the disk. At 14 orps, the dense spirals in the $t_{cool} = 1$ disk still somewhat trace the pattern in the $t_{cool} = 2$ disk, but they appear more open, penetrating further in and reaching further out. The $t_{cool} = 1/4$ disk, on the other hand, has considerably changed its structure by 14 orps. At 18 orps, the density distributions of the $t_{cool} = 1$ and 2 disks are different for all density ranges.

The rings develop at the same radii as in the $t_{cool} = 2$ simulation, again enhanced by shorter cooling times. The innermost rings form first and, since $Q \gg 2$ at $r < 10$ AU, they never become unstable and break into spirals. The possible exception is the 6.7 AU ring of the $t_{cool} = 1$ disk at 18 orps with $Q = 1.6$. However, inside and outside this ring, Fourier analyses for $t_{cool} = 1$ and 1/4 reveal ILR's for patterns with CR's at or beyond 10 AU, with the number and strength increasing as $t_{cool}$ diminishes. The point at which inward and outward mass transport separate is between 26 and 29 AU for all the simulations, as shown in Figure 10. Fourier analysis shows that this is the corotation radius for $m = 2$ (two-armed spiral) modes, which are the dominant patterns in $t_{cool} = 2$ and 1. The overall expansion rates observed in the $t_{cool} = 2$ and 1 disks seem to indicate that disk expansion rate goes as $1/t_{cool}$, but the duration of the $t_{cool} = 1/4$ is not sufficient to confirm this relationship. A small loss of angular momentum ($\approx 1.5\%$) is observed in these simulations, which can be accounted for by the mass that leaves the grid at the top and outer boundaries.

TABLE 2
Characteristics of the Constant $t_{cool}$ Disks

| Simulation | $t$ (orps) | $Q$ | $\dot{M}$ ($M_\odot$ yr$^{-1}$) | $\Sigma$ power | $T_{eff}$ power | $R_{disk}$ (AU) | $U_{tot}$ (erg) | Rings | Fragments |
|---|---|---|---|---|---|---|---|---|---|
| $t_{cool} = 2$ | 23.5 | 1.44 | 5(-7) | -2.50 | -0.71 | 52 | 9.4(40) | Yes | No |
| $t_{cool} = 1$ | 18.0 | 1.43 | 1(-6) | -2.47 | -0.73 | 60 | 9.9(40) | Yes | No |
| $t_{cool} = 1/4$ | 14.0 | 1.50 | 4(-6) | -2.48 | -0.65 | 63 | 8.4(40) | Yes | Yes* |

* Under high azimuthal resolution, see § 3.6.

Table 2 and Figure 13 compare various quantities of the constant cooling time runs. Table 2 shows the final values of several parameters for the three main simulations: the time in orps at which the simulation was stopped, the average Toomre $Q$ in the unstable zone (12 to 40 AU), the average mass transport rate both inward and outward after 12 orps, the power-law indices of the surface density and effective temperature, the outer radius of the disk, the final value for the total internal energy, and whether the disk develops rings and fragments. For this table, the outer disk radius is taken as the radius at which the power law of the surface density breaks down. Note that the average mass transport rates, about $5\times10^{-7}$, $10^{-6}$, and $4\times10^{-6}$ $M_\odot$ yr$^{-1}$ for $t_{cool} = 2$, 1, and 1/4, respectively, are inversely proportional to the cooling time. On the other hand, the asymptotic internal energy and average $Q$ are remarkably close for all three simulations. Outside the ring region, the azimuthally averaged $\Sigma(r)$'s follow roughly the same $r^{-5/2}$ power law for all three cases. Evidently, these characteristics of the asymptotic state do not depend strongly on the strength of cooling.

Figure 13 illustrates mass per cylindrical radius interval, surface density, and $Q$ as a function of radius for 14 and 18 orps for the three simulations. The inner rings and the $t_{cool} = 1/4$ void can be easily identified in the cylindrical mass plots. Notice how fast the disk grows as a function of cooling time in the surface density plots. The $r^{-5/2}$ power law extends much further out in the $t_{cool} = 1/4$ disk already at 14 orps. The $t_{cool} = 1$ simulation requires 4 orps to achieve a somewhat smaller size. It is obvious from the $Q$ plots how irregular this quantity is for $t_{cool} = 1/4$ compared to the other two disks even though the averages are similar. In general, $Q$ is higher between 20 and 30 AU in the $t_{cool} = 1/4$ case.

3.6. The $t_{cool} = 1/4$ High-Resolution Simulation

Tests performed in Paper I indicate that insufficient azimuthal resolution in the 3-D potential solver tends to smear gravitationally bound objects azimuthally and possibly prevents material from clumping, especially at large $r$ where the cylindrical cells become quite elongated (e.g., at 40 AU $r\Delta\phi \approx 2$ AU, while at all radii $\Delta r = 1/6$ AU). To test the effect of azimuthal resolution on clump formation, one last simulation is carried out with quadrupled azimuthal resolution. The radial and vertical resolutions are kept the same; however, the grid is truncated in the $r$ and $z$ directions, i.e., $(r,\phi,z) = (256,512,32)$, to reduce running time. Anything outside 42.3 AU in $r$ and 5 AU in $z$ is ignored in this calculation. The disk is gravitationally unstable between roughly 10 and 50 AU, and only low-density material populates the upper layers. So, if bound objects were to form, they would do so in the equatorial plane of the inner disk. A cooling time of 1/4 orp is used, and the disk is evolved from 12 to 14 orps.

By 12.2 orps, clumps which are $\geq 4$ orders of magnitude denser than their surroundings start to appear throughout the unstable part of the disk (Figure 14). The majority form where spiral arms meet. These only show up as elongated structures in the lower resolution simulation. As in most of the simulations in Paper I, no dense clump survives longer than a local orbit. It is not obvious what the principal mechanism for clump dispersal is in our simulations; but we plan to investigate this further in the future. We suspect that tidal, thermal, and shear stresses all play important roles. By 14 orps, no clumps are left.

# 4. DISCUSSION

The $t_{cool}$ = 2 simulation shows four clear phases, namely, the initial *axisymmetric* phase (0 to ~ 5 orps), the short but violent *burst* phase (~ 5 to ~ 6 orps), the *readjustment* phase (~ 6 to ~ 12 orps), and the *asymptotic* nonaxisymmetric phase (~ 12 to 23.5 orps). The axisymmetric phase is characterized by slow contraction and density build-up as the disk cools. Once the outer disk reaches $Q < 1$, a violent expansion of the outer disk and large-scale redistribution of mass occur as a result of global spiral modes that grow in about a rotation period. Comparisons with Rice et al. (2003) and Lodato & Rice (2004) (see also Pickett et al. 1998 and references therein) suggest that the occurrence of a burst may depend on initial conditions, as well as a constant $t_{cool}$ as a functon of radius. Just after the burst, the material in the strong spirals falls back, the disk becomes hot, and the amplitude of the GI's decreases. During the asymptotic phase, the balancing of cooling and heating processes eventually brings the disk to quasi-equilibrium, with an overall slow expansion of the outer disk combined with inflow of mass through the 10 to 29 AU region. When different cooling times are applied to the asymptotic phase, the nonaxisymmetric dynamic behavior becomes stronger in response to increased cooling, but the thermal characteristics of the quasi-equilibrium state remain similar.

## 4.1. Mass Transport

Little has been done in the past to study long-term evolution of gravitationally unstable disks in 3-D, but the early stages of unstable evolution are relatively well understood. Analyses by Pickett, Durisen, & Link (1997), Pickett et al. (1998), and Laughlin, Korchagin, & Adams (1998) show that the bulk of the mass and angular momentum transport happens on a short time scale, during the nonlinear growth and saturation of the spiral structure after the disk first becomes unstable. This is clearly observed in the $t_{cool}$ = 2 simulation, with both inward and outward mass flow at rates as high as $10^{-5}$ M$_\odot$ yr$^{-1}$ during the burst phase. After the readjustment phase, during which the disk slowly achieves quasi-equilibrium, the GI's achieve a roughly constant average amplitude, but with large spatial and temporal fluctuations. Mass and angular momentum are still transported by GI's, the disk continues to expand, and mass continues to move inward and outward from the CR's of the dominant underlying two-armed patterns. Inside corotation, orbiting material loses angular momentum and moves inward; outside corotation, gas gains angular momentum and expands outward.

As shown in Table 2, the average mass transport rates are inversely proportional to the cooling time. On the other hand, Figure 3 shows that mass transport rates are not steady at any radius, and mass moves both inward and outward. Therefore, it is not possible to assign a single turbulent viscosity $\alpha$ parameter that describes transport in the entire disk. Although we recognize that equation (2) does not strictly apply for non-steady accretion, we can use it to estimate the effective $\alpha$, call it $\alpha_{eff}$, that would produce the same average $\dot{M}$ over the 10 to 29 AU region. Evaluated at 20 AU, we get $\alpha_{eff} \approx 0.065$. According to Gammie's (2001) local treatment (see also Pringle 1981), predicted values of $\alpha$ as a function of $t_{cool}\Omega$ can be calculated using his equation (20), i.e.,

$$\alpha = \left[\Gamma(\Gamma-1)\frac{9}{4}t_{cool}\Omega\right]^{-1}, \qquad (5)$$

where $\Gamma = 3 - 2/\gamma$. For a constant $t_{cool}$ and a nearly Keplerian disk the quantity $t_{cool}\Omega$ is a function only of radius. When we apply equation (2) to measured transport rates at different $r$'s, our $\alpha_{eff}$'s are generally one or two orders of magnitude larger than predicted by equation (5). So, applying local arguments, like those behind equation (5), to GI's in our simulations underestimates the mass inflow substantially. This is not surprising because we find that global two-armed modes dominate transport. In our simulations, with an overall constant cooling time, the GI's are an intrinsically global phenomenon that cannot be captured or well approximated by an effective $\alpha$ (Laughlin & Różyczka 1996; Balbus & Papaloizou 1999). According to Lodato & Rice (2004), transport in disks is local phenomenon as long as $H(r)/r << 0.1$. As the $M_{disk}/M_*$ ratio increases, $H(r)/r \leq 0.1$, and global transport becomes more important. For the simulations presented here $0.04 < H(r)/r < 0.06$, which satisfy the above condition only marginally, so transport is more likely to depend on global conditions. We note that the magnitude of the mass transport rates in our Table 2 are comparable to those reported for other GI simulations of solar-system-sized disks (e.g., Boss 2002; Nelson et al. 2000).

### 4.2. Fragmentation

Fragmentation of the disk into dense clumps or arclets occurs when the cooling time in an unstable disk is small enough that heating associated with GI dynamics cannot be produced fast enough to counterbalance the energy loss rate (Tomley et al. 1991, 1994; Gammie 2001). According to Gammie's (2001) local, thin-disk treatment, this occurs when $t_{cool}\Omega \leq 3$, i.e., when the cooling time is less than about half a rotation period ($P_{rot} = 2\pi/\Omega$). As shown in Figure 15, this criterion is only met at the very outer disk, outside 79 and 54 AU for $t_{cool} = 2$ and $t_{cool} = 1$, respectively. For the $t_{cool} = 1/4$, fragmentation should occur in most of the disk, outside 21 AU in this case, if the result from Gammie (2001) applies.

The spiral structure in the low azimuthal resolution (128 zones) $t_{cool} = 1/4$ simulation is extremely thin and dense. Given the results from the potential solver tests performed in Paper I, however, this resolution is probably insufficient to allow the disk to fragment. As shown in Figure 14, with quadruple the azimuthal resolution (512 zones), the disk does fragment almost immediately, in about 0.2 orps, and a handful of clumps forms outside ~ 20 AU as predicted by the Gammie criterion. It appears that the essential features of disk fragmentation are captured by local, thin disk treatments. It is an open question whether the $t_{cool} = 2$ and 1 cases would have undergone fragmentation with higher resolution, though it seems unlikely given Figure 15, except perhaps in the very outermost regions. Rice et al. (2003) performed 3-D SPH simulations of disks with $t_{cool}\Omega$ = constant and confirmed that their disks fragment only if $t_{cool}\Omega \leq 3$. They also observed a tendency for the value of $t_{cool}\Omega$ at which fragmentation sets in to increase slightly as $M_{disk}/M_*$ increases. Johnson & Gammie (2003) explored this problem further by adding a treatment of radiative cooling with realistic opacities to Gammie's local, thin disk calculations and found fragmentation to be strongly dependent on sharp changes in the opacities, in the sense that the initial cooling time may not be a reliable predictor of fragmentation. Gammie's criterion appears to be more a rule of thumb than a universal law.

### 4.3. Ring Formation

The formation of rings in the inner disk during the asymptotic phase is the most unexpected result of the simulations presented here and in Paper I. These rings form in the hot (Q > 2), GI stable inner (r < 10 AU) region, and they accumulate mass as it is brought inward from the unstable part of the disk. As explained in more detail in another paper (Durisen et al. 2004), analysis of both axisymmetric and nonaxisymmetric waves in and near the ring region indicates that ILR's due to spiral modes with CR's in the GI active region, combined with dissipation of wave energy, are responsible for ring growth. The growth rate of the rings seems to be inversely proportional to the cooling rate, although this relationship is hard to determine for the $t_{cool}$ = 1/4 disk.

Other simulations in Paper I and new simulations by our group with radiative cooling (Mejía 2004) also show ring-like structures where there is a gravitationally unstable zone adjacent to an inner stable zone, a plausible configuration for real protostellar disks (e.g., Gammie 1996; Armitage, Livio, & Pringle 2001). In addition to borders between low and high *Q* regions, similar rings might appear in GI active disks at other natural boundaries in disk structure, for example at drastic changes in opacity (e.g., Johnson & Gammie 2003) or in the equation of state of the gas. Durisen et al. (2004) point out that such rings are sites where gas giant planet formation could occur due to 1) an enhancement of the local gas surface density and 2) the rapid drift of solids to the peak of the rings (Haghighipour & Boss 2003a, 2003b). In the first case, the rings may themselves become gravitationally unstable, producing Jupiter-sized clumps in circular orbits (Paper I; Pickett & Lim 2004), which are shielded from the violent stresses of the GI-active region that destroy clumps in our simulations. On the other hand, the rapid drift of solid bodies to ring centers could substantially shorten core accretion (e.g., Pollack et al. 1996) time scales and ultimately lead to rapid planet formation (Pickett & Lim 2004; Durisen et al. 2004). Opacity transitions in and near the rings due to growth of solids may also enhance this process. GI-assisted core accretion model is a hybrid scenario that could solve the problem of rapid gas giant planet formation even if GIs or core-accretion do not efficiently form planets as separate mechanisms.

To our knowledge, the formation of multiple concentric rings in the gravitationally stable part of a GI-active disk has not been reported in numerical experiments by other researchers. Reasons may have to do with choices of gas equation of state (e.g., isothermal in Boss 2000; Mayer et al. 2004), rapid cooling (e.g., Boss 2001, 2002, 2004), or use of $t_{cool}\Omega$ = constant rather than $t_{cool}$ = constant (e.g., Rice et al. 2003; Lodato & Rice 2004). All of these effects tend to make GI's more nearly equally strong throughout a disk and might therefore suppress the formation of an active/inactive boundary. Our new simulations with realistic radiative cooling (Mejía 2004) suggest that $t_{cool}$ = constant and slow cooling may better represent the behavior of real disks than the other cases just mentioned.

4.4. Energy Balance

In the asymptotic phase, the Toomre $Q$ distribution of the unstable zone, between 12 and 40 AU, is more or less constant with time for all cooling times. This is consistent with the feedback loop between cooling and heating processes as proposed by Goldreich & Lynden-Bell (1965), Lin & Pringle (1987, 1990), and Bertin (1997). The 2-D N-body simulations of Tomley et al. (1991, 1994) suggested that the asymptotic $Q$ at a given radius would be lower for lower cooling time. We do not confirm this. In fact, the average asymptotic value of $Q$ seems relatively independent of the cooling time and averages about 1.43 to 1.50 in the unstable zone for all $t_{cool}$.

Figure 9 indicates that the highest temperatures in a vertical column of the disk occur away from the equatorial plane. Pickett et al. (2000a, 2000b) explain that the shortest relative heating times, compared to the local rotation period, occur in the high-altitude, low-density layers of the disk. The fact that midplane temperatures are lower than upper-layer temperatures also means that our disks are convectively stable, unlike those presented by Boss (2002, 2004). We observe vertical motions due to the wave dynamics of GI's, but these motions are not convective eddies.

4.5. Implications for Real Disks

The mass transport rates observed during the burst ($10^{-5}$ $M_\odot$ yr$^{-1}$) and the asymptotic phases ($5\times10^{-7}$ $M_\odot$ yr$^{-1}$) for $t_{cool}$ = 2 are consistent with accretion rates during FU Orionis outbursts and during early phases of T Tauri disk evolution, respectively (Hartmann & Kenyon 1985, 1987a, 1987b; Bell et al. 1995, 2000; Gullbring et al. 1998). The nature of the burst is probably sensitive to details of the disk initial conditions. However, any disk that evolves from a stable to an unstable state by cooling or by build up of a dead zone (Armitage et al. 2001) should experience a similar event. This can lead to detectable outbursts comparable in magnitude to FU

Ori if the mass transport rates are as high as the ones observed in the $t_{cool} = 2$ case and continue all the way to the star. The inward mass motion reported in §§3.3.1 and 3.4.1 does not penetrate the inner disk because these simulations do not have turbulent transport to carry mass through the hot GI-stable inner disk. The duration of the burst ($\approx 200$ yr) is about twice as long as the maximum lifetime of a single FU Ori outburst (Bell et al. 1995), but, again, burst duration probably depends on our pre-determined cooling time and initial conditions.

The transport rates in the asymptotic phase are sustained by gravitational instabilities, and the average stays approximately constant during the duration of this phase ($\approx 3000$ yr) in the $t_{cool} = 2$ simulation. These mass transport rates will probably only diminish over much longer times as the disk evolves. If the disk's overall growth rate ($5.6 \times 10^{-3}$ AU yr$^{-1}$) holds for a long period of time, it could reach a size of 300 - 400 AU in radius in about $5 \times 10^4$ yr. Mass inflow at a rate of $5 \times 10^{-7}$ M$_\odot$ yr$^{-1}$ in a disk with $M_{disk} = 0.07$ M$_\odot$ cannot be sustained for more than about $10^5$ yr, the typical transition age between FU Orionis and T Tauri systems. It is likely that both the mass transport and the expansion will decrease on this or a shorter time scale until the disk reaches more typical average T Tauri size and accretion rates.

The surface density of all these disks is roughly proportional to $r^{-5/2}$, too steep according to what most authors believe are the surface density profiles ($\propto r^{-3/2}$) of non-irradiated disks (see for example D'Alessio et al. 1998, Figure 7). However, the effective temperature distribution as a function of radius ($T_{eff} \propto r^{-0.65}$ to $T_{eff} \propto r^{-0.73}$ between 3 and 50 AU) is close to the power laws that best fit the observations of such disks (D'Alessio et al. 1998, Figure 3). These simulations used a very simple cooling prescription, not radiative cooling, and so they were not expected to reproduce observational features in detail.

## 5. SUMMARY

We presented the longest-to-date, global, 3-D, grid-based hydrodynamics simulations of marginally stable protoplanetary disks that reach instability by cooling. We have confirmed that disks relax to a quasi-equilibrium asymptotic state with a constant average $Q$ as a result of balance between cooling and heating, independent of the cooling time applied. Mass transport rates and outer disk expansion rates in the asymptotic state are inversely proportional to the cooling time. Mass transport is dominated by low-order global spiral modes of the disk operating over 10's of AU, and it is much stronger than estimated by a local, thin disk treatment and is not well represented by a simple effective $\alpha$. From these simulations, it is clear that gravitational instabilities are an important mechanism for angular momentum and mass transport, that they can produce realistic T Tauri mass accretion rates, and that their onset could be the cause of FU Orionis events. The disk fragmentation criterion proposed by Gammie (2001) from local, thin disk simulations does seems to apply accurately to our global 3-D simulations. Fragmentation is only possible numerically with high azimuthal resolution, which was not used in all cases. None of the dense clumps produced survives for more than one orbit, and all clumps are gone in our short cooling time, high-resolution calculation within two orps of first clump formation. Each simulation presented here resulted in the production of thin, dense rings of material in the inner, hotter part of the disk. It remains to be determined whether these rings can become gravitationally unstable and produce bound companion objects or serve as reservoirs for rapid planet formation by the core accretion mechanism.

## ACKNOWLEDGEMENTS

The authors thank F.C. Adams, A.P. Boss, P. Cassen, S.A.E.G. Falle, C.J. Gammie, T.W. Hartquist, G. Lodato, L. Mayer, A. Nelson, J. Papaloizou, T.R. Quinn, W.K.M. Rice, and D. Whittet for useful discussions and insight while preparing this manuscript. The work presented in this paper was supported by NASA Grant NAG5-11964 from the Origins of Solar Systems Program and by NASA Grant NAG5-10262 from the Planetary Geology and Geophysics Program. This work was also supported in part by systems obtained by Indiana University through its relationship with Sun Microsystems Inc. as a Sun Center of Excellence. MKP also thanks the hospitality of the Department of Physics and Astronomy at the University of Leeds, where she has been a Visiting Scholar. ACM acknowledges the NASA Astrobiology Institute for support.

REFERENCES

Adams, F.C., & Lin, D.N.C. 1993, in Protostars and Planets III, ed. E. Levy & J. Lunine (Tucson: University of Arizona Press), 721

Adams, F.C., Ruden, S.P., & Shu, F.H. 1989, ApJ, 347, 959

Armitage, P.J. 1998, ApJ, 501, L189

Armitage, P.J., Livio, M., & Pringle, J.E. 2001, MNRAS, 324, 705

Balbus, S.A., & Hawley, J.F. 1991, ApJ, 376, 214

Balbus, S.A., & Hawley, J.F. 1997, in ASP Conf. Ser. 121, Accretion Phenomena and Related Outflows, IAU Colloq. 163, ed. D. Wickramasinghe, G. Bicknell, & L. Ferrario (San Francisco: ASP), 90

Balbus, S.A., & Hawley, J.F. 2000, Space Sci. Rev, 92, 39

Balbus, S.A., & Papaloizou, J.C.B. 1999, ApJ, 521, 650

Bate, M.R. 1998, ApJ, 508, L95

Bell, K.R., Lin, D.N.C., Hartmann, L., & Kenyon, S.J. 1995, ApJ, 444, 376

Bell, K. R., Cassen, P. M., Wasson, J. T., & Woolum, D. S. 2000, in Protostars and Planets IV, ed. V. Mannings & S. Russell (Tucson: University of Arizona Press), 897

Bertin, G. 1997, ApJ, 478, L71

Boss, A.P. 1997, Science, 276, 1836

Boss, A.P. 1998, ApJ, 503, 923

Boss, A.P. 2000, ApJ, 536, L101

Boss, A.P. 2001, ApJ, 563, 367

Boss, A.P. 2002, ApJ, 576, 462

Boss, A.P. 2003, LPS, 34, 1054

Boss, A.P. 2004, LPS, 35, 1124

Cassen, P.M., Smith, B.F., Miller, R.H., & Reynolds, R.T. 1981, Icarus, 48, 377

Christodoulou, D. 1991, Ph.D. dissertation, Louisiana State University

D'Alessio, P. 1996, Ph.D. dissertation, Universidad Nacional Autonóma de México

D'Alessio, P., Cantó, J., Calvet, N., & Lizano, S. 1998, ApJ, 500, 411

Durisen, R.H., Gingold, R.A., Tohline, J.E., & Boss, A.P. 1986, ApJ, 305, 281

Durisen, R.H., Mejía, A.C., Pickett, B.K., & Hartquist, T.W. 2001, ApJ, 563, L157

Durisen, R.H., Mejía, A.C., & Pickett, B.K. 2003, RRDAp, 1, 173

Durisen, R.H., Cai, K., Mejía, A.C., & Pickett, M.K. 2004, Icarus, submitted

Gammie, C.F. 1996, ApJ, 457, 355

Gammie, C.F. 2001, ApJ, 553, 174

Goldreich, P., & Lynden-Bell, D. 1965, MNRAS, 130, 97

Gullbring, E., Hartmann, L., Briceño, C., & Calvet, N. 1998, ApJ, 492, 323

Haghighipour, N., & Boss. A.P. 2003a, ApJ, 583, 996

Haghighipour, N., & Boss. A.P. 2003b, ApJ, 598, 1301

Hartmann, L., & Kenyon, S.J. 1985, ApJ, 299, 462

Hartmann, L., & Kenyon, S.J. 1987a, ApJ, 312, 243

Hartmann, L., & Kenyon, S.J. 1987b, ApJ, 322, 393

Hawley, J.F., Gammie, C.F., & Balbus, S.A. 1996, ApJ, 464, 690

Imamura, J.N., Durisen, R.H., & Pickett, B.K. 2000, ApJ, 221, 937

Johnson, B.M., & Gammie, C.F. 2003, ApJ, 597,131

Klahr, H.H., & Bodenheimer, P. 2003, ApJ, 582, 869

Larson, R. 1984, MNRAS, 206, 197

Laughlin, G., & Bodenheimer, P. 1994, ApJ, 436, 335

Laughlin, G., & Różyczka, M. 1996, ApJ, 456, 279

Laughlin, G., Korchagin, V., & Adams, F.C. 1998, ApJ, 504, 945

Lin, D.N.C., & Papaloizou, J.C.B. 1996, ARA&A, 34, 703

Lin, D.N.C., & Pringle, J.E. 1987, MNRAS, 225, 607

Lin, D.N.C., & Pringle, J.E. 1990, ApJ, 358, 515

Lodato, G., & Rice, W.K.M. 2004, MNRAS, 351, 630

Mantos, M.A., & Cox, D.P. 1998, ApJ, 509, 703

Mayer, L., Quinn, T., Wadsley, J., & Stadel, J., 2002, Science, 298, 1756

Mayer, L., Quinn, T., Wadsley, J., & Stadel, J., 2004, ApJ, submitted

Mejía, A.C. 2004, Ph.D. dissertation, Indiana University

Miyama, S.M., Hayashi, C., & Narita, S. 1984, ApJ, 279, 621

Morfill, G.E., Spruit, H., & Levy, E.H. 1993, in Protostars and Planets III, ed. E. Levy & J. Lunine, (Tucson: University of Arizona Press), 939

Nelson, A.F., Benz, W., Adams, F.C., & Arnett, W. 1998, ApJ, 502, 342

Nelson, A.F., Benz, W., & Ruzmaikina, T.V. 2000, ApJ, 529, 357

Norman, M.L., & Winkler, K.-H.A. 1986, in NATO Advanced Workshop on Astrophysical Radiation Hydrodynamics, ed. K.-H.A. Winkler & M.L. Norman (Dordrecht: D. Reidel)

Paczyński, B. 1978, AcA, 28, 91

Papaloizou, J.C.B., & Lin, D.N.C. 1995, ARA&A, 33, 505

Papaloizou, J.C.B., & Savonije, G.J. 1991, MNRAS, 248, 353

Pickett, B.K. 1995, Ph.D. dissertation, Indiana University

Pickett, B.K., Cassen, P.M., Durisen, R.H., & Link, R. 1998, ApJ, 504, 468

Pickett, B.K., Cassen, P.M., Durisen, R.H., & Link, R. 2000a, ApJ, 529, 1034

Pickett, B.K., Durisen, R.H., Cassen, P.M., & Mejía, A.C. 2000b, ApJ, 540, L95

Pickett, B.K., Durisen, R.H., & Davis, G.A. 1996, ApJ, 458, 714

Pickett, B.K., Durisen, R.H., & Link, R. 1997, Icarus, 126, 243

Pickett, B.K., Mejía, A.C., Durisen, R.H., Cassen, P.M., Berry, D.K., & Link, R. 2003, ApJ, 590, 1060 (Paper I)

Pickett, M.K., & Lim, A.J. 2004, A&G, 45, 12

Pringle, J.E. 1981, ARA&A, 19, 137

Pollack, J.B., O. Hubickyj, O., Bodenheimer, P., Lissauer, J.J., Podolak, M., & Greenzweig, Y., 1996, Icarus 124, 62.

Rice, W.K.M., Armitage, P.J., Bate, M.R., & Bonnell, I.A. 2003, MNRAS, 339, 1025

Shakura, N.I., & Sunyaev, R.A. 1973, A&A, 24, 337

Stone, J.M., Gammie, C.F., Balbus, S.A., & Hawley, J.F. 2000, in Protostars and Planets IV, ed. V. Mannings, A. Boss & S. Russell (Tucson: University of Arizona Press), 589

Tomley, L., Cassen, P.M., & Steiman-Cameron, T.Y. 1991, ApJ, 382, 530

Tomley, L., Steiman-Cameron, T.Y., & Cassen, P.M. 1994, ApJ, 422, 850

Toomre, A. 1964, ApJ, 139, 1217

van Albada, G.D., van Leer, B., & Roberts, W.W.Jr. 1982, A&A, 108, 76

Velikhov, E. 1959, SovPhysJETP, 36, 995

Williams, H. 1988, Ph.D. dissertation, Louisiana State University

Yang, S. 1992, Ph.D. dissertation, Indiana University

Yorke, H.W., & Bodenheimer, P. 1999, ApJ, 525, 33

FIGURE CAPTIONS[1]

Figure 1: Properties of the initial axisymmetric disk. The top panel shows meridional contours of number density. The bottom panel shows the radial distribution of the Toomre $Q$, the midplane temperature $T_{mid}$ in units of 10 K, and the midplane number density $n_{mid}$ in units of $10^{11}$ cm$^{-3}$. $Q$ is calculated using the full vertical (top + bottom) surface density, and both the sound speed and omega (replacing the epicyclic frequency for a nearly Keplerian disk) at the equatorial plane.

Figure 2: Evolution of the $t_{cool}$ = 2 orps disk. All the images show number densities at the equatorial plane using the same grayscale (colorscale in the electronic version) as Figure 1. Each square is 170 AU on a side, which enclose the disk at maximum size. The top section shows the entire evolution every 2 orps; the bottom section shows the onset of the instabilities between 4 and 7 orps.

Figure 3: Average mass transport rate of $t_{cool}$ = 2 for two different times. Shown here are the $\dot{M}$'s calculated by differences in the total mass fraction as a function of radius between 12 and 18 ORPs, and between 12 and 23.5 ORPs. The grey thin lines are the maximum and minimum values for the instantaneous transport rate during the 12 - 23.5 ORP time interval.

Figure 4: Rings in the inner disk. a) The inner disk at 22 orps, when the third ring becomes most pronounced. Each side of the square is 85 AU. b) Total mass per grid cylinder in solar masses at various times during the evolution. The cylindrical shells used to measure $M_{cyl}$ are 1/6 AU wide.

Figure 5: Periodogram plot of periods detected for two-armed ($m$ = 2) patterns in the $t_{cool}$ = 2 evolution. The dashed lines show the locations of the rings. The solid horizontal line (29 AU) divides regions of net inflow and net outflow over long periods of times. The curves from top down are the OLR, CR and ILR for $m$ = 2.

Figure 6: Total internal energy $U_{tot}$ as a function of time. The luminosity of the disk is obtained by dividing the internal energy by the cooling time, therefore it follows the same curve.

Figure 7: Toomre $Q$ at 0, 6, 12, 18, and 23.5 orps vs. radius. The curves roughly overlap between 12 and 40 AU after 12 orps.

---

[1] Figures and Mpeg animations available at http://westworld.astro.indiana.edu

Figure 8: Comparison of the $t_{cool}$ = 2 disk at 6 orps (top four panels) and at 23.5 orps (bottom four panels). Clockwise from upper left: midplane number density, midplane temperature, effective temperature, and heating time. The contours in the heating panels illustrate the density structure for comparison. The unshaded regions show where the number density is lower than $\approx 10^9$ cm$^{-3}$ and no thermal physics is calculated.

Figure 9: Comparison of the $t_{cool}$ = 2 disk at 6 orps (top) and at 23.5 orps (bottom) for one slice of the grid (left) and the azimuthal average of all the slices (right). Each box measures 50 AU along the radial direction and 5 AU in the vertical direction, zooming in on the inner disk. From top to bottom: number density, temperature, and heating time. As in Figure 8, number density maps show cells below the limiting density for heating and cooling physics of $\approx 10^9$ cm$^{-3}$. The rest of the maps show only where the cooling and heating are applied.

Figure 10: Final mass transport rates for all three simulations vs. radius.

Figure 11: Total internal energy of all the $t_{cool}$ simulations as a function of time. See Figure 6.

Figure 12: Comparison of the constant cooling time disks at 12, 14, and 18 orps.

Figure 13: Comparison of various quantities for the three constant $t_{cool}$ runs. The first row shows mass per grid cylinder vs. radius at 14 and 18 orps. The second and third rows show similar plots of the surface density and the Toomre $Q$. See text for further explanation.

Figure 14: Comparison between the low- and high-resolution $t_{cool}$ = 1/4 simulations. The top images show a zoom in on the inner 42 AU of the $t_{cool}$ = 1/4 disk. The bottom images show the High-res disk at the same times and with the same scale.

Figure 15: $t_{cool}\Omega$ for various cooling times vs. radius. The dashed line indicates $t_{cool}\Omega$ = 3.